BMC
Bioinformatics

                                    **Open Access**

# Intuitive representation of surface properties of biomolecules using BioBlender

Raluca Mihaela Andrei[1,2], Marco Callieri[3], Maria Francesca Zini[1,5], Tiziana Loni[1], Giuseppe Maraziti[4], Mike Chen Pan[1,6], Monica Zoppè[1]*

## Abstract

**Background:** In living cells, proteins are in continuous motion and interaction with the surrounding medium and/or other proteins and ligands. These interactions are mediated by protein features such as electrostatic and lipophilic potentials. The availability of protein structures enables the study of their surfaces and surface characteristics, based on atomic contribution. Traditionally, these properties are calculated by physico-chemical programs and visualized as range of colors that vary according to the tool used and imposes the necessity of a legend to decrypt it. The use of color to encode both characteristics makes the simultaneous visualization almost impossible, requiring these features to be visualized in different images. In this work, we describe a novel and intuitive code for the simultaneous visualization of these properties.

**Methods:** Recent advances in 3D animation and rendering software have not yet been exploited for the representation of biomolecules in an intuitive, animated form. For our purpose we use Blender, an open-source, free, cross-platform application used professionally for 3D work.
On the basis Blender, we developed BioBlender, dedicated to biological work: elaboration of protein motion with simultaneous visualization of their chemical and physical features.
Electrostatic and lipophilic potentials are calculated using physico-chemical software and scripts, organized and accessed through BioBlender interface.

**Results:** A new visual code is introduced for molecular lipophilic potential: a range of optical features going from smooth-shiny for hydrophobic regions to rough-dull for hydrophilic ones. Electrostatic potential is represented as animated line particles that flow along field lines, proportional to the total charge of the protein.

**Conclusions:** Our system permits visualization of molecular features and, in the case of moving proteins, their continuous perception, calculated for each conformation during motion. Using real world tactile/sight feelings, the nanoscale world of proteins becomes more understandable, familiar to our everyday life, making it easier to introduce "un-seen" phenomena (concepts) such as hydropathy or charges. Moreover, this representation contributes to gain insight into molecular functions by drawing viewer's attention to the most active regions of the protein. The program, available for Windows, Linux and MacOS, can be downloaded freely from the dedicated website http://www.bioblender.eu

* Correspondence: mzoppe@ifc.cnr.it
[1]Scientific Visualization Unit, Institute of Clinical Physiology, CNR of Italy, Area della Ricerca, Pisa, 56124, Italy
Full list of author information is available at the end of the article

**BioMed** Central

## Background

The fact that we humans are very good at extracting information through visual observation is well synthesized in the old adage "a picture is worth a thousand words". The solution of the 3D structure of myoglobin in 1958 by Kendrew [1] marked the beginning of the new era of structural biology. Since then, a wealth of protein structures has been solved and today the Protein Data Bank (PDB) counts over 67.000 protein structures [2,3].

With the availability of all these data, and with the advance of computer graphics (CG) technologies, tools for the visualization of 3D structures were created such as VMD [4,5], SPDBViewer [6,7], Chimera [8,9], PyMOL [10,11] and others. Balls and sticks for atoms and bonds, ribbons for the secondary structures, and molecular surfaces are some of the possible representations of proteins. Most programs can also calculate surface features such as electrostatic potential (calculated with APBS [12] or DelPhi [13]) and hydropathy (Kyte-Doolittle [14]). When present, these features are represented as field lines and/or as ranges of colors.

Since the late '90s, the development of CG techniques has advanced at spectacular pace. Among the most widely used tools, is the art and science of 3D animation. This technique consists in the creation and animation of 3D objects (complete with surfaces, skeletons, and simulated physical properties) in a virtual world, which can be 'filmed' using virtual cameras and lights. Several programs are available for this, including the commercial packages Maya/Autodesk, 3D Studio Max and Softimage XSI (all from Autodesk, [15]), Cinema 4D (from MAXON Computer GmbH [16]) and the open-source Blender [17].

Not surprising, all of these have been used for the study and representation of biological molecules and processes. Some examples are collected and visible on dedicated websites [18,19]. The films range from the simple representation of the mechanical functioning of a single protein, to complex events involving many subjects such as DNA replication and RNA processing, to views of major cellular processes, such as apoptosis, etc.. These latter ones are important scientific efforts and add to their educational value the bonus of rising interest in the general public to approach biology.

For our purpose we use Blender, an open-source, free, cross-platform 3D application. Blender is a powerful instrument for 3D modeling, animation, gaming and rendering, that provides a complete workbench for producing still images, simple animations or very complex scenes with thousand of objects in motion, all textured, lighted and filmed for proper view.

Traditionally, the process of creating a 3D animation film consists of a number of steps roughly grouped in modeling, animation, rendering, special effects and compositing. Blender offers a platform to elaborate and integrate all of these steps. Objects are created in the virtual world by modeling them in the 3D scene starting from primitives or by importing them from other programs. A *time line* holding key frames (points in time in which objects have defined configuration set-ups) is used to animate the objects in the scene in various ways: by direct rotations/translations of the object, by mesh deformation obtained moving its components (vertices, edges, faces), via skeleton (inverse or forward kinematics) or by using the Game Engine (GE), typically deployed in video games. Additionally, physics-based animations can be achieved by simulated forces such as gravity, magnetic, vortex, wind *etc*. Objects are given a surface appearance by the use of material shaders and textures. These two elements define the behavior of the surface when illuminated, by specifying local information like color, reflectance (dull or shiny) and microstructure (roughness or smoothness).

Once the animation and texturing is defined, the scene is equipped with other assets such as a background, lights and cameras and the process concludes with the 'filming' (rendering of all frames which are assembled to generate a video).

In this article, we illustrate a step forward in the direction of using bio-animation both as a divulgation and as a discovery tool. Our aim is to visualize molecules in a directly informative way, also showing their motion obtained from structural data (Figure 1). This task is done using BioBlender [20], in which Blender is used to access several scientific programs. BioBlender is an engine built in Blender with an interface for biological visualization (Figure 2).

The use of Blender GE to elaborate the movement of proteins, starting from 2 or more conformations is described in Zini *et al.* (BioBlender: Fast and Efficient All Atom Morphing of Proteins Using Blender Game Engine, manuscript submitted). Briefly, starting from data from NMR collections or X-rays of the same protein crystallized in different conditions, we use Blender GE, equipped with special rules approximately simulating atomic behavior, to interpolate between known conformations and obtain a physically plausible sequence of intermediate conformations. This sequence is output as a list of pseudo .pdb file (list of atoms with their x, y, z coordinates) which are the basis for the visual elaboration described here.

As the result of this study, we propose a new visual code for the representation of two important surface properties: electrostatic potential (EP) and molecular lipophilic potential (MLP). Using features different from color permits their simultaneous delivery in photo-realistic images leaving the utilization of color space for the description of other biochemical information. Here we describe the details of this process.

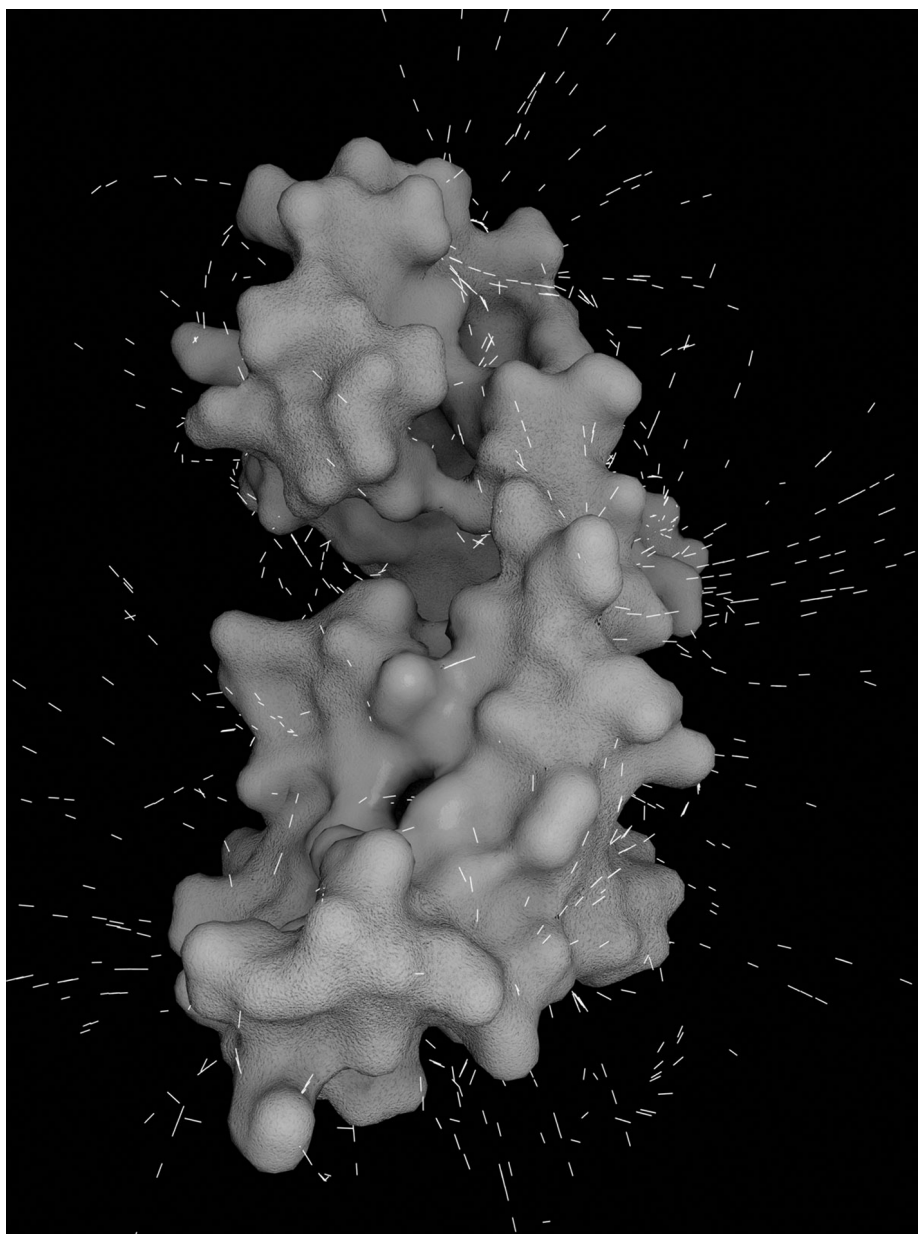

**Figure 1 Example of BioBlender representation**. The protein (Calmodulin) is shown with its chemical and physical features represented according to the proposed code, as described in the present article. The image is a single frame from an animated sequence, showing EP and MLP. For a 3D interactive example, please visit http://www.scivis.ifc.cnr.it/images/stories/3d_interactive/VIS_CaCaM/VIS_CaCaM.html.

## Methods

### Programs and scripts

**BioBlender** is an extension of Blender, in which custom Python scripts have been implemented for building the interface, importing the meshes and the curves, converting MLP values into vertex colors and managing various scientific programs (http://www.bioblender.eu).

In the construction of BioBlender, we have made ample use of several existing programs, listed here.

**Blender 2.5** is a free, open source, cross platform suite of tools for 3D creation [17].

**PyMOL 1.2r3pre** is a Python-enhanced molecular graphics tool [10], used for visualization of .pdb files. It calculates the electrostatic potential through APBS plugin. This tool is also used to generate the 3D mesh of the molecular surface for the molecule. The obtained geometry is exported in .wrl format, easily read by 3D software tools.

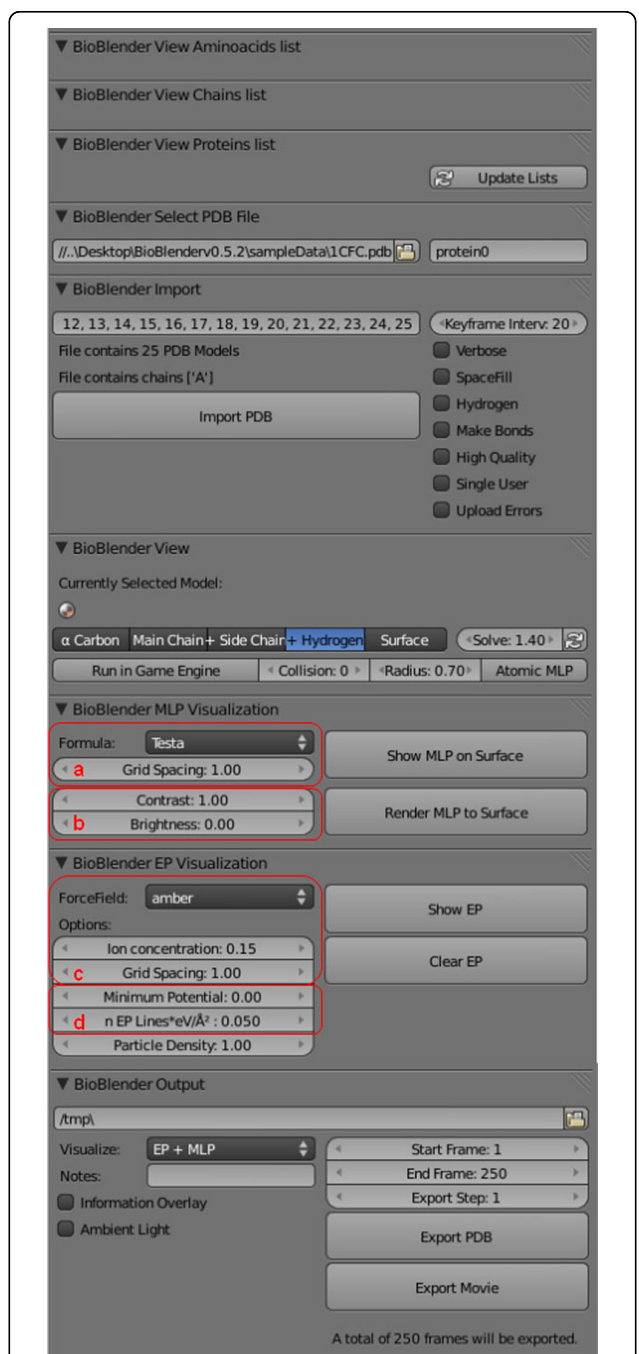

**Figure 2 BioBlender interface**. The interface is structured in 9 panels: *amino acids list* - to select and highlight amino acids in the 3D viewport, *chains list* - to select different protein chains, *proteins list* - to select different proteins; *select .pdb file* - upload from user defined path, or access directly from PDB.org specifying the 4 letter code; *import* - at the import phase, it is possible to select various parameters, including covalent/Van der Waals radius, include/exclude Hydrogens and others; *view* - visualization in 3D working space, activation of Game Engine; *MLP visualization* - Parameters for MLP; *EP visualization* - parameters for EP; *output* - export of .pdb files and rendered frames. a: choice of formula and grid spacing; b: contrast and brightness control; c and d: calculation and representation, respectively.

**PDB2PQR-1.6.0** [21,22] is a software package that automates many of the common tasks of preparing structures for continuum electrostatics calculations, providing a platform-independent utility for converting protein files from PDB format to PQR format. It assigns partial atomic charge to every atom according to different force fields (AMBER 94, CHARMM 27 or PARSE) and saves a .pqr file in which the occupancy and temperature columns are replaced by atomic charge and radius, respectively. It also adds missing hydrogens, calculates pKa values and generates an input (.in) for APBS calculations. The .in file stores information on the 3D dimension of the protein, the ionic concentration of solvent, biomolecular and solvent dielectric constants. Ionic concentration of 0.150 mol/l NaCl, biomolecular dielectric constant of 2 and solvent dielectric constant of 78.54 (water) were used for our calculation.

**APBS-1.2.1 (Adaptive Poisson-Boltzmann Solver)** [12] is a software for evaluating the electrostatic properties of nanoscale biomolecular systems, through solution of the Poisson-Boltzmann equation. APBS takes as inputs a .pqr and an .in file and calculates the electrostatic potential in every point of a grid in the protein space, which is output as a .dx file.

**scivis.exe** is a custom software written in C++ used to calculate the field lines and to export them in a ASCII file to be imported in Blender. This tool imports the 3D surface (.obj) and the EP grid (.dx) calculated by APBS. The computation of the field lines is a multi-step process: EP values are mapped on the 3D surface, a gradient field is calculated in the volume containing the molecule, an automatic selection of areas with high values of EP is done and the corresponding field lines are computed for these areas using the gradient field. When used as primary application, in addition to the described features, scivis.exe provides visual feedback for all its processing steps. It is possible to visualize the molecular surface, the EP grid, the gradient grid and the field lines.

**Python 2.6** is an interpreted, interactive, object-oriented, extensible programming language [23]. In this project, Python has been used in different stages, both as a scripting component of various software tools (like Blender and PyMOL) and as a stand-alone scripting language.

**pyMLP.py** is a Python script written and kindly provided by Julien Lefeuvre (available from [24]); it contains a library of atomic lipophilic potential for all atoms present in proteins (we added the values for some monosaccharides and nucleic acids) and it calculates the MLP in every point of a grid in the protein space according to various formulae such as Fauchere, Dubost, Brasseur, etc. (we introduced Testa formula). The grid step can be changed by the user to cope with the protein size and computer performances (in terms of memory occupancy and calculation time).

## Results

We present here a software/method to produce the simultaneous visualization of EP and MLP on proteins. In the case of moving proteins, the program produces a rendered animation, in which every second of the resulting movie contains 25 images (24-30 frames per second is the standard video speed), and at every frame the shape, EP and MLP of the molecule are automatically recalculated.

In the elaboration of each frame representing proteins, still or in motion, the steps of object (mesh) creation, surface calculation and data manipulation for EP and MLP are elaborated independently using both scientific and CG programs to obtain the series of frames compositing the animation (Figures 3 and 4).

### Protein surfaces

The molecular surface of proteins [25] is calculated in PyMOL starting from the .pdb file, as shown in Figure 3, upper left. For series of conformations (obtained with Game Engine or derived from molecular dynamics), the procedure is reiterated. PyMOL was chosen because the surfaces created by this software have a regular triangulation even at low polygon resolution and it is only marginally afflicted by the problem of internal disjoint surfaces. In the 3D mesh used in the example reported in Figure 4A and in other tests with wider range of dimensions (number of polygons between 4.5 and 50 thousands), all the triangles have similar areas. The mesh is exported by PyMOL as a .wrl, a file which contains information about the position of the vertices, edges, characteristics of the material of the polygon *etc.*

### MLP calculus

The MLP calculus (Figure 3 upper right) is done using pyMLP.py [26,27]. This script calculates the lipophilic potential in every point of a grid in the space of the protein and exports the values in a .dx file. The script contains a library of atomic lipophilic potential values for every atom based on its chemistry, and several formulae

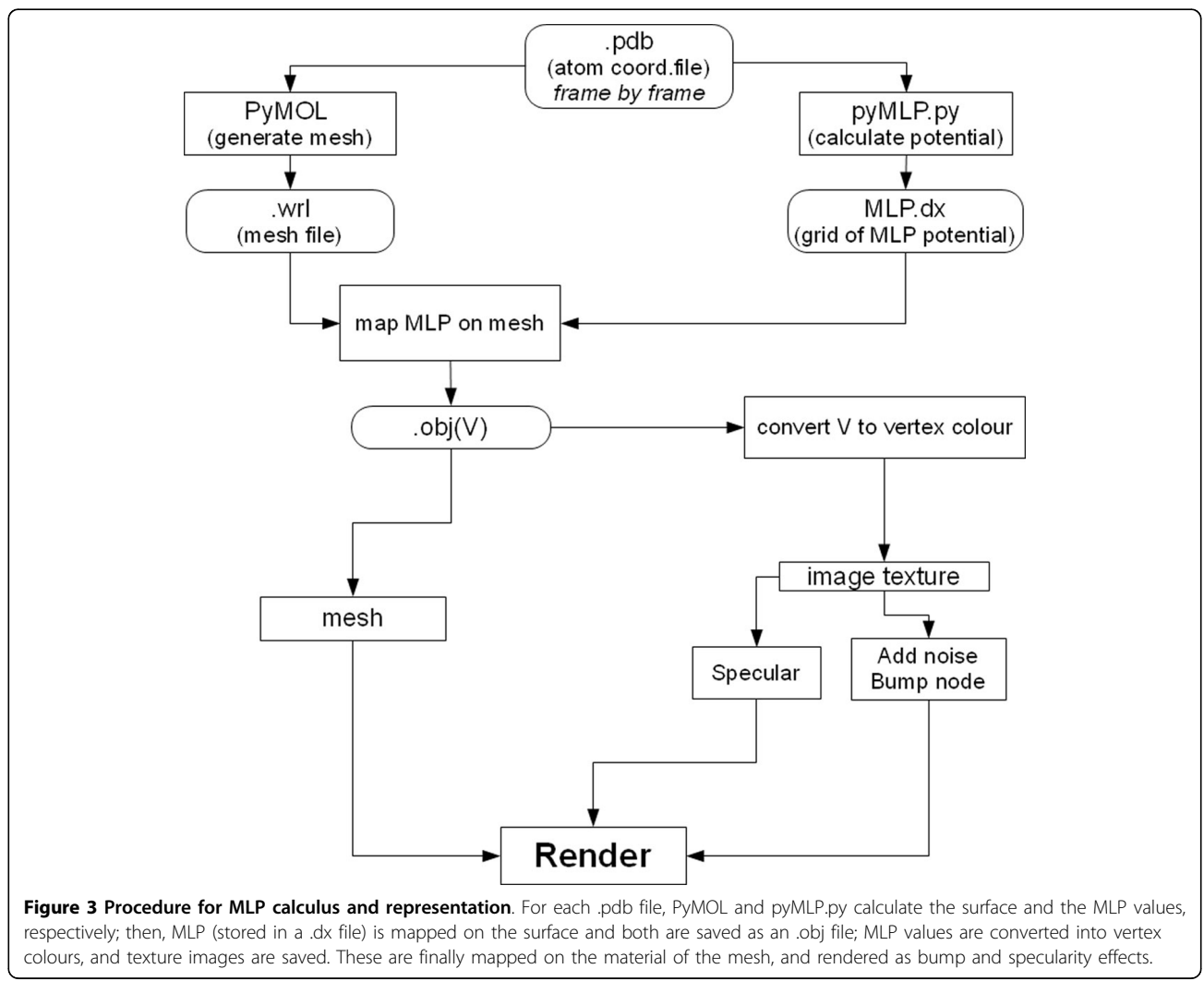

**Figure 3 Procedure for MLP calculus and representation**. For each .pdb file, PyMOL and pyMLP.py calculate the surface and the MLP values, respectively; then, MLP (stored in a .dx file) is mapped on the surface and both are saved as an .obj file; MLP values are converted into vertex colours, and texture images are saved. These are finally mapped on the material of the mesh, and rendered as bump and specularity effects.

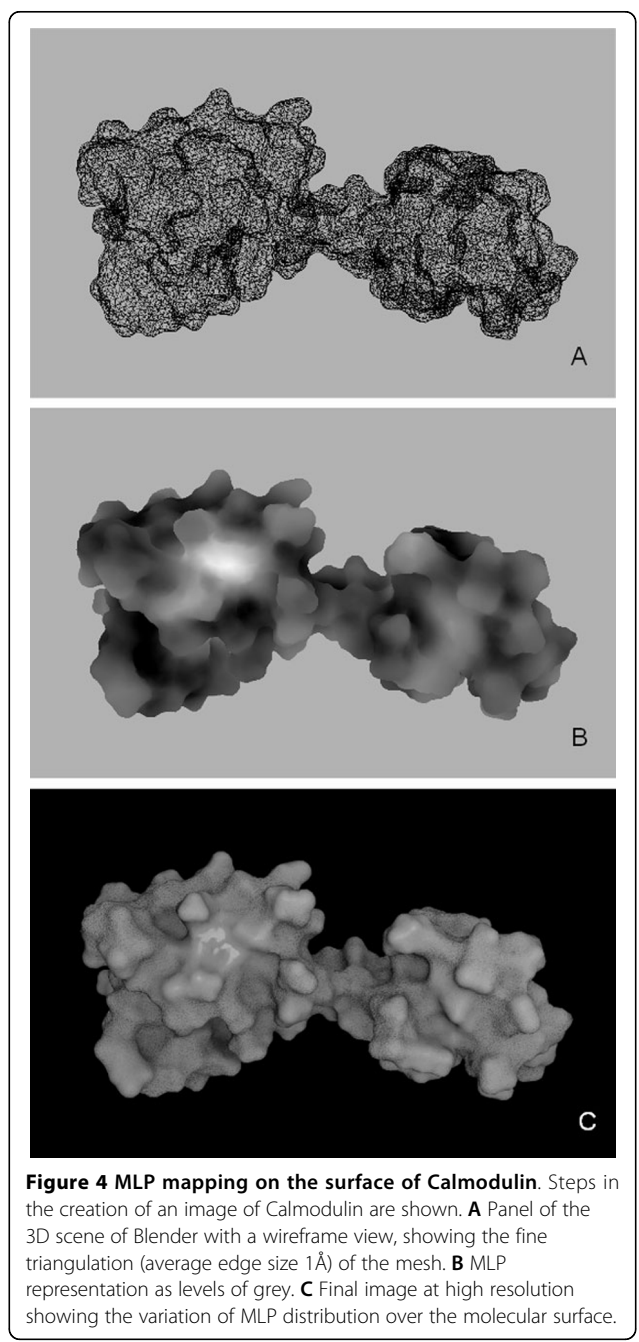

**Figure 4 MLP mapping on the surface of Calmodulin**. Steps in the creation of an image of Calmodulin are shown. **A** Panel of the 3D scene of Blender with a wireframe view, showing the fine triangulation (average edge size 1Å) of the mesh. **B** MLP representation as levels of grey. **C** Final image at high resolution showing the variation of MLP distribution over the molecular surface.

for MLP calculation. However it does not support the Testa formula [28],

$$MLP(r) = \sum_i f_i e^{\frac{-|r-r_i|}{2}}$$

where r is a point in the protein space, $f_i$ is the atomic lipophilic potential for atom i, and $r_i$ is the position of atom i.

The Testa formula is an atom-based function that uses the Broto [26] fragment scheme and an exponential distance function, appropriate for protein calculations; therefore we modified pyMLP.py to include Testa formula. The MLP accuracy depends on the grid spacing (a in Figure 2); in BioBlender the default is set at 1Å, a dimension comparable to the mean size of the triangle edge of the 3D mesh; this parameter is a good compromise between MLP data, mesh triangulation, computer memory and time for calculation (see below).

pyMLP outputs a .dx file in which the header defines the grid origin, the grid step and the number of points on each axis.

**MLP rendering**
The code for the representation of hydropathy that we propose is a range of optical features that go from smooth-shiny surface (hydrophobic) to rough-dull (hydrophilic), as shown in Figure 4C.

Data elaboration for rendering is done in a series of steps (Figure 3, lower part):

*1. MLP values mapping on the mesh.* The MLP values (typically between -3 and 1 for soluble, membrane-embedded and cytoplasmic proteins) are mapped on the surface of the molecule by assigning values of MLP to the mesh. The algorithm (included in a custom program, OBJCreator) is simple: for every vertex of the mesh, the correspondent grid-cell, in the MLP grid, is identified and the value of potential is calculated using trilinear interpolation and assigned to the vertex.

This process is very fast (about 2 seconds on a personal computer for calmodulin) and the mesh vertex density is high enough to represent smoothly the potential spatial transition. The information about the MLP values corresponding to every vertex are stored in the V field of an .obj file as texture coordinates (U and V).

*2. MLP values conversion into vertex colors.* MLP values (previously assigned to the vertices of the mesh) are converted into vertex colors, assigning the same value for each RGB channel, to obtain levels of gray). For the conversion we normalize the range of the MLP values ([-3,1]) to the range of gray scale ([0,1]), and set value 0 of MLP to correspond to the value 0.5 of the gray scale. In this way the hydropathy of the protein is visualized in Blender as levels of gray: bright areas representing hydrophobicity and dark areas hydrophilicity (Figure 4B). The use of the default conversion scale provides a coherent representation for all proteins; however, at this step, to enhance MLP features for any particular protein under study, the user can modify contrast and brightness using sliders (b in Figure 2).

*3. Creation of the first image texture.* The mesh is unwrapped to generate a texture parametrization and the per-vertex color values are saved ('baked') in a texture image. UV unwrapping is a procedure that consists in flattening a 3D object (e.g. the world globe) on a 2D plane (e.g. the world map), so that each vertex of the 3D mesh is assigned a correspondent 2D texture coordinate [29]. The 2D image is also called image texture or UV map, where U and V are the texture axes.

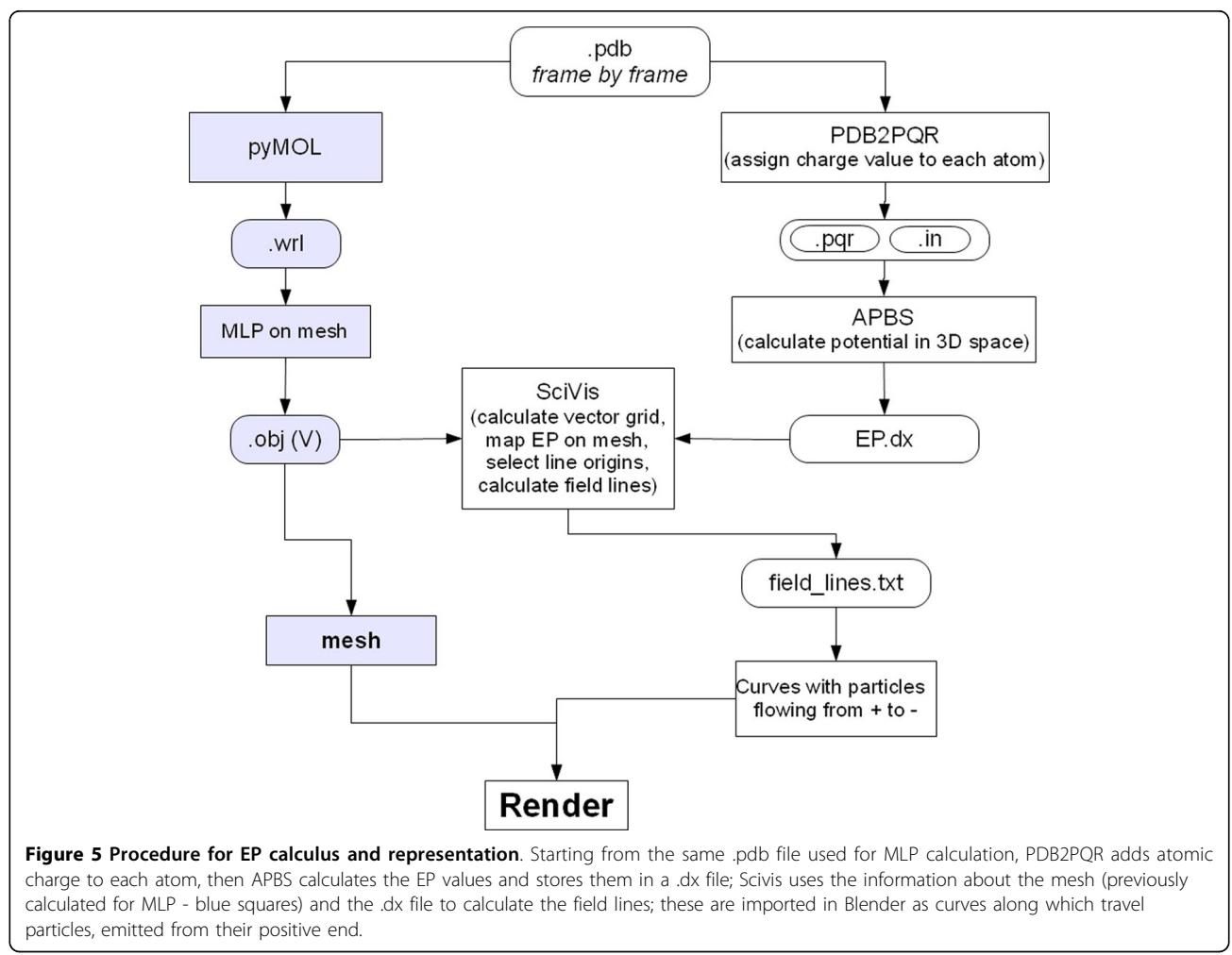

**Figure 5 Procedure for EP calculus and representation**. Starting from the same .pdb file used for MLP calculation, PDB2PQR adds atomic charge to each atom, then APBS calculates the EP values and stores them in a .dx file; Scivis uses the information about the mesh (previously calculated for MLP - blue squares) and the .dx file to calculate the field lines; these are imported in Blender as curves along which travel particles, emitted from their positive end.

*4. Creation of the second image texture*. In order to make the more hydrophilic areas rough the procedure involves the addition of a noise pattern of amplitude proportional to the degree of gray of the texture. This is achieved using the Node Editor of Blender, adding a Gaussian noise to the texture image, which produces an image with a strong noise over the black regions, gradually reduced on gray regions to reach a level of no noise on white. In the rendering process this noise is converted to bump, as explained below.

*5. Addition of specularity and roughness*. In the final rendering step, the image obtained in the first step (gray scale) is finally mapped on specularity from dull to shiny, and the second image is mapped on bump. Bump mapping is a rendering technique generally used to represent very small scale geometry like scratches, roughness or graininess. This technique does not affect the geometry of the object: the perceived local geometry is only an optical effect obtained by light reflection modifications. In the final image hydrophobic areas are represented as reflective and smooth, while the more hydrophilic ones as duller and rougher (Figure 4C). By avoiding the use of color, as well as of gray scale, the differences in color are only due to the effect of light interacting with the surface, i.e. the darker areas are the least illuminated.

## EP calculus

While the use of movies is mostly intended to show transition between conformations of a protein, it also allows the introduction of special effects of CG to convey other information. We have elaborated the following procedure using both BioBlender and external programs to display the EP associated with molecular (partial) charges (see Figure 5, right side). All programs are accessed through BioBlender interface, also used to set specific parameters.

The .pdb file used for mesh creation and MLP calculus is submitted to PDB2PQR program [21,22] which outputs 2 files: .pqr and .in. These files store information on the size and charge of every atom, and on the dimensions of the protein, the ionic concentration, biomolecular and solvent dielectric constant, respectively. Both .pqr and .in are input files for APBS program [12], that calculates the electrostatic potential in every point of a grid in the space of the protein and exports the values in a .dx file, analogous to the one seen above for MLP. The force field, the ion concentration and the grid spacing can be set by the user (c in Figure 2).

EP is redrawn as field lines calculated by a custom software, scivis.exe, that combines information from the mesh file (.obj) with EP values (see below). This computation comprises different steps:

*1. Mapping EP on the surface mesh*
*2. Transformation of the grid of EP values into a grid of gradients*
*3. Selection of more active surface areas by weighted Monte Carlo sampling*
*4. Drawing of field lines that are stored in a .txt file*

The EP values are mapped on the surface of the protein by assigning a value of EP to every vertex of the mesh, as seen above for MLP, i.e. trilinear interpolation.

A grid of gradient vectors is built starting from the scalar field of EP values: for each point the gradient is calculated according to the values in neighbor points, finding the direction and slope of EP change.

The gradient data are used to generate the field lines in the space surrounding the protein. From the infinite possible lines, we are interested in generating a 'meaningful' subset associating the lines associated with areas of the mesh with high value of EP, obtaining a distribution of lines that is proportional to the surface EP value: more lines will rise in the more electrically active areas, and the total number of lines will be proportional to the global level of potential of the molecule. This selection is done by Monte Carlo sampling weighted with respect to the potential value of the surface in each area.

For the selection of this subset, the user has two controls (d in Figure 2): the absolute EP value on the surface from which the creation of the field lines starts (lines are generated only in areas with an EP higher than a threshold - Minimum potential) and a parameter that represents the general line density (expressed as Number of lines × eV/Å$^2$). By modulating this parameter users can select the most appropriate value for a group of proteins, obtaining a concentration of field lines which is coherent across the various proteins.

Once the 'interesting' locations (points) are selected, the lines are calculated by following the gradient in both directions, iteratively moving with small steps according to the gradient (small-step integration). Line points are added until one of the following three conditions is met: 1. the limit of the calculated grid is reached, 2. the line intersects the mesh or 3. the field is too low (the gradient is approximately 0 or equal to the value set by the user). The lines are saved as sequences of points in an ASCII file (.txt).

Thanks to the random nature of the selection procedure, lines do change every time the procedure is run but the more electrically active areas (where more lines are present) are readily identifiable. This property proves to be particularly effective when represented in animation, since it gives the idea of fuzziness, useful for electricity representation, while conveying the information about EP distribution on the surface.

In the case of Calmodulin, depicted in Figure 1, and even more evident in the WebGL animated representation, most lines are directed towards the surface, due to the fact that the protein is slightly acidic, with an isoelectric point of 4.09.

**EP representation**

Field lines are imported into Blender as NURBS curves which are not rendered (they are invisible in the final image), but instead are used to guide a particle effect. Every curve starts at its most positive end which is associated with a particle emitter. The particles, drawn as short segments, flow along the curves from positive to negative, respecting the field lines convention in physics. During animation, the particles are generated every 5 frames (0.2 sec) and have a life-time of 20 frames (0.8 sec). This means that the system is in steady state after the sixteenth frame (see the scheme in Figure 6). Representation of EP as moving particles on a trajectory, played in time, is interpreted easily and transmits the idea of polarity of the charged areas of a biomolecule.

If the user is interested in visualization of only one conformation, the animated particles are displayed/played in loop (they are emitted for 250 frames and have a lifetime of 20 frames).

**Moving proteins**

In the visualization of proteins in motion, every frame is elaborated as a single .pdb file. Because at every frame the atomic coordinates change, also the surface features (shape itself, EP and MLP, calculated by integrating the atomic values) change accordingly, and must be recalculated. Due to extremely high-level modifications (topology changes, merging/separation of surface parts) it is not possible to use a single geometry and animate it through conventional tools. It is instead necessary to rebuild the surface geometry, importing a new set of mesh coordinates at each frame. This implies a very large amount of calculations, but allows the elaboration of a sequence of images that is coherent from frame to frame, thus giving the impression of continuity.

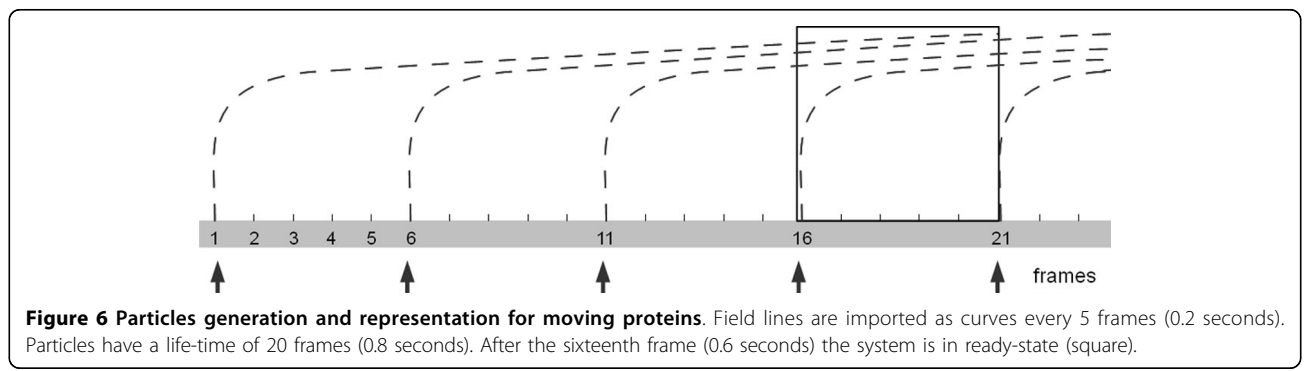

**Figure 6 Particles generation and representation for moving proteins.** Field lines are imported as curves every 5 frames (0.2 seconds). Particles have a life-time of 20 frames (0.8 seconds). After the sixteenth frame (0.6 seconds) the system is in ready-state (square).

In summary, for each frame (conformation) we visualize MLP as textured mesh and EP as curves and animated particles. The result is a sequence of frames showing the moving protein with its properties, EP and MLP, represented together: MLP as a range of visual and tactile characteristics and EP as flow of particles that move from positive to negative along invisible field lines (as shown in the movie Protein Expressions - Study N. 3 [30]).

## Discussion

The description of biological phenomena has always made use of graphical presentation, starting from the early botanical and zoological drawings, including famous anatomical folios, that greatly help viewers, professionals and not, to understand and learn about nature.

Since the early times, an artistic component has been included, often unnoticed by viewers, but greatly exploited by the scientists/artists. Even today, the clearest graphical descriptions of natural and artificial subjects are hand-or CG-drawn rather than photographic images. The 'artistic' dimension allows for a better interpretation of the subject, the choice of illumination, and the removal of irrelevant details and disturbing effects.

The same attitude has motivated a number of scientists to use various graphical tricks when showing data related to structural features of macromolecules. Although most structural information contained in a .pdb file (a list of atoms and their 3D coordinates) is actually 'readable', biologists typically use graphical programs to explore protein structures; indeed the literature has an abundance of such programs, including some very popular. These programs can transfer the structural information from a linear list of atoms to a 3D virtual space and display it on 2D surface; positional information is interpreted with the aid of chemical information stored in libraries (of amino acids, nucleotides and other molecules), that introduce chemical bonds, electric charges, hydrophobicity scales and so on. In this way the user is enabled to observe features of the molecules of interest according to her/his needs. Recent years have seen the development of 3D computer graphics techniques that have culminated in the recent success of the blockbuster movie Avatar, in which an entire world has been created in CG, including 'floating mountains' and forest with thousands of (CG built!) plants, animals, insects *etc.*

Similar techniques can be used to show the nanoscopic world of cells, populated with all sorts of environments, proteins, nucleic acids, membranes, small molecules and complexes. Indeed, there are several remarkable examples of efforts in this new discipline of Bio-Animation, some of which have reached a large public. Beside the beauty and the educational value of these animations, we consider that the very process of creating such movies includes a heuristic importance both in the development of the graphical instruments and in the studies implied in the elaboration of the subjects' (proteins) movements and interactions. In fact, when a researcher is induced to take a different point of view, such as needed for the visual elaboration, s/he will be exposed to possible new insight, facilitating better understanding of the process under study. In this way a novel spatial reasoning can complement the classical biochemical reasoning typically employed in molecular research.

Our group is among those involved in the development of animated biology, and in this paper we report one aspect of such effort, namely the elaboration, using Blender, of a code capable of showing two of the most critical features that determine the behavior of macromolecules: their electrostatic and lipophilic potentials.

### Choice of Blender

Among the professional packages developed for CG, one only has the double advantage of being open source and available free of charge: Blender.

Blender is the result of a world-wide, concerted effort to put tools of the highest standard for CG creations at the reach of any artist (or scientist) regardless of her/his capability of paying for such tools. The project is guided by the non-profit Blender Foundation, and animated by countless developers that voluntarily devote time and effort to constantly introduce the most up to date techniques into the package, equipping users with any instrument they need. Blender 2.5, the latest major release, introduced a new design of the user interface, new physics engine for smoke (volumetric), particles and soft bodies, and, importantly, the possibility to achieve all Blender's functions from scripting, through APIs.

## BioBlender

On the framework of Blender 2.5, we built BioBlender, which includes a section specifically built for biological work. Inside BioBlender, for the analysis of proteins structures, various types of visualization are available: alpha carbon, main chain, main chain and side chains, all atoms (including hydrogens) and molecular surface. The elaboration of proteins' motions and the simultaneous representation of surface physico-chemical properties of proteins in motion are the innovations that BioBlender introduces in macromolecular visualization.

## Elaboration of protein motion

We use Blender Game Engine to elaborate the movement of proteins, when more than one conformational state is known. Starting from data from NMR collections or X-ray of the same protein crystallized in different conditions, we use Blender GE, equipped with special rules approximately simulating atomic behavior, to interpolate between known conformations and obtain a physically plausible sequence of intermediate conformations. This sequence can be explored within Blender or can be output as a list of pseudo .pdb file (list of atoms and x, y, z coordinates) which are the basis for the visual elaboration.

It is important to notice that this procedure can be applied to any .pdb or (better) sequence of .pdb files representing a continuous series describing a conformational transition, obtained by Blender or by any other means, e.g. Molecular Dynamics simulation.

## Visualization of moving proteins, and of their molecular surface features

The development of structural biology that made available tens of thousands of structures, not only improved our knowledge on structural features such as the richness of protein folds (secondary and tertiary structure), and of their association in groups (quaternary structure). It also increased knowledge associated with protein motion: in fact most proteins exert their function through some kind of motion. This is best understood by observing the movement in an animated film. The role of side chains, which are the determinants of such motions, is at present difficult to appreciate by using present visualization tools that either provide a fixed all-atom structure, or show dynamically only a limited number of atoms.

We present here a procedure that allows the direct observation of moving proteins focusing on their surface features, rather than on their structure. In particular, we have focused on hydropathy and electrical fields as they appear on and around the molecular surface.

These features can be calculated and visualized by a number of programs, which typically display them with a color code. We reasoned that for these properties a more 'photo-realistic' display would help viewers in the de-codification of their meaning, and elaborated the system here reported. Example of the use of these codes can be seen for a single protein in the Proteopedia page [31] (see also Additional file 1) and for a complex in our movie Protein Expressions - Study N3 [30].

The main idea of the proposed visual mapping is to exploit perceptual associations between the values to be mapped and visual characterization of real-world objects. Ideally, by using already established perceptual association, the viewer will be able to understand the provided information more naturally, without the use of explicit legends. For MLP mapping, two opposite surface characterizations able to convey a sense of affinity to water or to oil were selected. In our real-world experience, a very smooth, hard surface (like porcelain or wax) is completely impervious to water but can be easily coated by oil. The opposite visual feedback is associated to grainy, crumbly, dull surfaces (like clay bricks or biscuits) which can be easily imagined being soaked in water. These considerations led to the 'painting' of highly hydrophobic areas as shiny, smooth material and of highly hydrophilic areas as dull and rough.

While the MLP value is only observable on the surface itself, electrical phenomena are associated to the idea of an effect projected in the volume surrounding a charged object, and able to affect other objects (like the high school favorite amber rod attracting paper bits). Field lines are a common way to describe the effect of the electrical field. EP value is therefore represented by showing small particles, moving along the path defined by field lines, visualizing a high concentration of particles in areas where the electrical field is stronger.

The representation of both features in black and white allows the viewer to grasp their values, without distracting with arbitrary information which is not interpretable if not associated with a de-coding legend, making it easier to interpret.

For MLP elaboration we considered that none of the available programs are accurate enough to provide useful information: most molecular displaying packages simply attribute a fixed value of MLP to every atom of a given amino acid, using the Kyte-Doolittle scale [14]. This scale was elaborated almost 30 years ago with the aim of identifying structural features of proteins, namely the interior portions of globular proteins and membrane spanning segments in membrane associated proteins, but is not indicated for the evaluation of the distribution of MLP on the molecular surface. Indeed, some other programs include a more appropriate method of calculation, such as VASCo [32] which employs the Brickman [33] formula on an atom based library and a Fermi-type distance function. We have implemented a calculation with the Testa formula, which uses an atom-based fragment scheme and an exponential function. The values thus obtained are plotted on the vertices of the molecular surface. This procedure results in a very smooth distribution of MLP values which is then displayed with a scale of 'tactile' textures, ranging from dull-rough to shiny-smooth. The advantage of such calculation and representation is mostly noticeable in animated movies showing the transition between different conformation of proteins, when patches of hydrophobic areas are gradually exposed on the surface of proteins which will facilitate docking onto other macromolecules.

For EP, we developed a visual code based on a flow of particles (small lines) flowing towards the negative pole: this is particularly useful for the observation of interacting molecules and for molecules whose field is changing when the conformation changes. To elaborate EP we made use of several programs and integrated them in a flow whose final result is the continuous display of the EP and its development during protein conformational transitions.

### Time considerations

The entire process is very fast: a protein of 2262 atoms is imported in 7 sec, while MLP and EP computation with grid spacing 1 Å take 70 and 19 seconds, respectively, on a standard personal PC equipped with Windows XP, Intel Core 2 Duo CPU, 2.33 GHz, 3.25 GB RAM.

Our example is Calmodulin: after activation due to the binding of 4 Calcium ions, the protein undergoes a major conformational transition in which both its EP and its MLP change considerably: the Ca ions introduced in the 4 EF hand domains affect the EP by virtue of their own charge and the MLP by inducing the opening of each globular domain to expose two major hydrophobic patches which enable the protein to interact with its partners and push the calcium signal downstream in the biochemical pathway.

Proteins and their surface properties can also be visualized in a 3D interactive way on web platform exploiting the new WebGL component of HTML5. Using this API, it is possible to display 3D content in a web page without the use of external plug-ins, by writing an appropriate visualization program using the OpenGL syntax. Using a javascript support library, SpiderGL [34], we built an interactive visualization scheme [35] which accepts as input the same data (meshes, field lines and the MLP texture) calculated by BioBlender.

### Conclusions

In conclusion, we have developed a computational instrument that allows the display of molecular surfaces of moving (or still) proteins, putting special emphasis on their electrical and lipophilic properties. We consider that this representation allows better (or at least more immediate and intuitive) understanding of the dynamical forces governing intermolecular interactions and thus facilitate new insights and discoveries.

### Additional material

> **Additional file 1: Calmodulin in motion.** The movie (in .avi format) shows several transitions of calmodulin in the Apo form (without Calcium) and the major conformational change induced by the binding of 4 Ca ions. The movie can also be seen online at http://proteopedia.org/wiki/index.php/Calmodulin#Calmodulin_in_Motion.

### Abbreviations

EP: Electrostatic Potential; MLP: Molecular Lipophilic Potential; GE: Game Engine; CG: Computer Graphics; 3D: three-dimensional.

### Acknowledgements

We thank the PDB2PQR, APBS, PyMol teams, late Warren DeLano and the Blender users and developers community for kind answers to our many questions. Work funded by Regione Toscana grant 'Animazione 3D' to MZ. This article has been published as part of *BMC Bioinformatics* Volume 13 Supplement 4, 2012: Italian Bioinformatics Society (BITS): Annual Meeting 2011. The full contents of the supplement are available online at http://www.biomedcentral.com/1471-2105/13/S4.

### Author details

¹Scientific Visualization Unit, Institute of Clinical Physiology, CNR of Italy, Area della Ricerca, Pisa, 56124, Italy. ²Scuola Normale Superiore, Pisa, 56125, Italy. ³Visual Computing Lab, ISTI, CNR of Italy, Area della Ricerca, Pisa, 56124, Italy. ⁴Big Bang Solutions, Navacchio (Pisa), 56023, Italy. ⁵University of Pisa, Pisa, 56125, Italy. ⁶Harvard Medical School, Boston, MA 02115. USA.

### Authors' contributions

RMA performed research, wrote and tested software; MC, MFZ, GM contributed programming; MC, MCP contributed scivis.exe and BioBlender interface, respectively; TL, MZ contributed visual elaboration with Blender; MZ conceived research; RMA, MZ wrote paper.

### Competing interests

The authors declare that they have no competing interests.

Published: 28 March 2012